\pdfoutput=1
\documentclass[12pt, a4paper]{article}
\usepackage[T1, T2A]{fontenc}
\usepackage[utf8]{inputenc}
\usepackage[english, russian]{babel}
\usepackage{amsmath, amssymb}
\usepackage[left=3cm, right=1.5cm, top=2cm, bottom=2cm]{geometry}
\usepackage[displaymath]{lineno}
\usepackage[space]{cite}
\usepackage{xcolor, graphicx}

\begin{document}

\newpage

\begin{center}
{\large\bf ION AND ELECTRON TEMPERATURE RECOVERY FOR IRKUTSK INCOHERENT SCATTER RADAR MEASUREMENTS} \\[10pt]
V. P. Tashlykov, A. V. Medvedev, V. P. Lebedev, and I. V. Kichigin
\end{center}

\noindent Recovering ion and electron plasma temperatures from Irkutsk Incoherent Scatter Radar (IISR) data is a challenge due to rapid Faraday rotation, which significantly distorts signal correlation profiles. This paper presents an adaptation of the full-profile analysis (FPA) method tailored to the specific conditions of the IISR using a two-step inverse problem framework. The core novelty of the approach lies in the integration of data from both narrowband and wideband channels: experimental power profiles from the wideband channel are incorporated directly into the forward model for temperature recovery. By mathematically establishing the link between the wideband signal amplitude and the temperature ratio, we introduce a non-linear constraint that significantly narrows the solution search space. This addresses the problem of "stretched" global minima in the objective functional, effectively filtering out physically inconsistent results. The inverse problem is solved using the Improved Stochastic Ranking Evolution Strategy (ISRES), which enables fully automated parameter estimation without the need for manual initial guesses.

\newpage

\section*{INTRODUCTION}

The incoherent scatter (IS) radio wave technique \cite{Beynon1978} is the most comprehensive ground-based tool for ionospheric plasma diagnostics, providing vertical profiles of electron density ($N_e$), electron ($T_e$) and ion ($T_i$) temperatures, ion drift velocity ($V_i$), ion composition, and other parameters. The theoretical foundation describing the spectral characteristics of the backscattered signal was laid in classical works \cite{Dougherty1960, Evans1969}. The high diagnostic capability of the technique relies on the close agreement between experimental data and theoretical spectra, which are non-linear functions of the plasma parameters. However, extracting these parameters from recorded radar echoes is a complex non-linear inverse problem, which has evolved from independent parameter estimation within individual range gates (gate analysis) to full-profile fitting across the entire altitude range \cite{Lehtinen1986, Holt1992, Hysell2008}.

Conventional range-gated analysis neglects vertical parameter correlations, which is justifiable only when using complex modulation schemes, such as alternating codes that provide high range-lag localization of the radar ambiguity function \cite{Lehtinen1986}. In the general case, to properly account for range and lag smearing, full-profile analysis (FPA) must be employed, which embeds the full two-dimensional radar ambiguity function directly into the forward model \cite{Holt1992, Lehtinen1986}. Further development of these principles within Bayesian filtering frameworks allows controlling parameter gradients both in altitude and time, yielding higher stability under low signal-to-noise ratio (SNR) conditions \cite{Virtanen2021, Ross2022}. However, such approaches remain computationally intensive, motivating the search for more efficient "near-optimal" processing algorithms \cite{Nikoukar2008}. A comprehensive review of hardware upgrades and IS data processing techniques can be found in \cite{Robinson2004}.

The specific operational features of the Irkutsk Incoherent Scatter Radar (IISR) \cite{Zherebtsov2002, Medvedev2009, Kushnarev2017} significantly complicate the application of standard FPA methods. The IISR operates in the 154--162~MHz frequency band using a single, strictly linearly polarized antenna, which induces deep Faraday power fading in the received signal \cite{Shpynev2004, Shpynev2017}. This phenomenon severely distorts correlation profiles, making the inverse temperature recovery problem for IISR even more ill-posed. Previously proposed algorithms \cite{Shpynev2004} relied heavily on expert judgment and required frequent operator intervention to adjust initial guesses, making them poorly suited for automated processing of continuous long-term datasets.

A major challenge in IISR temperature recovery is the appearance of physically unrealistic $T_i$ values, where the diurnal variation of ion temperature exhibits unphysically large swings (e.g., up to 1000~K at 250~km), contradicting independent observations. The primary hypothesis of this work is that daytime $T_i$ values are systematically overestimated due to the "stretched" ravine-like topology of the objective function's global minimum, which is typical for IS inverse problems \cite{Shpynev2010, Vallinkoski1988}. Under low SNR conditions, the optimization algorithm can become trapped in parameter domains that formally minimize the fitting residual but violate physical reality. To address this issue, we propose an automated computational framework incorporating a new system of a priori constraints tailored to IISR conditions. These constraints effectively stabilize the inverse problem, filtering out physically inconsistent solutions and ensuring reliable recovery of altitude-temporal temperature distributions.

\section{Methodology}
This section describes the developed method for recovering ionospheric plasma temperature profiles from IISR data, based on a two-step solution of the inverse problem. The proposed approach combines information from narrowband (pulse length ~700~$\mu\text{s}$) and wideband (total pulse length ~200~$\mu\text{s}$) channels, which allows compensating for distortions caused by rapid Faraday rotation and significantly stabilizing the search for the global optimum.

The methodology is presented in the following sequence:
\begin{itemize}
\item Subsections 1.1--1.3 present the mathematical formulation of the optimization problem and describe the physical models of signal formation and plasma scattering cross-section;
\item Subsection 1.4 considers the preliminary stage of wideband signal analysis to obtain the concentration profile and reference values;
\item Subsections 1.5 and 1.6 formulate the system of spline-based a priori constraints and the algorithmic strategy for finding the solution;
\item Subsection 1.7 describes the method for determining the error intervals of the obtained inverse problem solution;
\item Subsection 1.8 describes the physical approximations used in formulating the forward problem and estimates the temperature recovery errors;
\item Subsection 1.9 presents the final unified mathematical model describing the detailed formulation of the forward problem and the inverse problem solution strategy.
\end{itemize}

\subsection{Nonlinear Optimization Problem Formulation}
Let us formulate the problem of recovering ionospheric plasma temperature profiles $T_e$ and $T_i$ from IISR data as a classical nonlinear optimization problem. Mathematically, the problem reduces to finding the global minimum of the functional $\Psi(\mathbf{x})$:
\begin{equation}
\Psi(\mathbf{x}) = f(\mathbf{x}) + \lambda \Phi(g(\mathbf{x})),
\label{eq:objective_func}
\end{equation}
where the first term $f(\mathbf{x})$ represents the residual objective function for the least squares method:
\begin{equation}
f(\mathbf{x}) = \sum_{s,t} (y_{\text{model}}(\mathbf{x}, s, t) - y_{\text{exp}}(s, t))^2.
\end{equation}
Here $s$ and $t$ are space and time coordinates, $y_{\text{exp}}$ represents experimental measurements, and $y_{\text{model}}$ is the forward problem model.

The second term $\lambda \Phi(g(\mathbf{x}))$ accounts for a priori constraints given as a system of inequalities:
\begin{equation}
g_j(\mathbf{x}) \le 0, \quad j = 1, 2, \dots, m.
\end{equation}
The penalty function $\Phi(g(\mathbf{x}))$ is defined as follows [1]:
\begin{equation}
\Phi(g(\mathbf{x})) = \sum_{j=1}^{m} \max(0, g_j(\mathbf{x}))^2.
\end{equation}
The penalty value increases as the sought solution $\mathbf{x}$ approaches the boundaries of the feasible region, with the penalty parameter $\lambda$ determined heuristically. Unlike local optimization algorithms (such as COBYLA), which require a sufficiently accurate initial guess, nature-inspired algorithms (NIA) \cite{Kumar2023} claiming global convergence are often used to minimize this functional. In Section 1.6, we consider one such algorithm adapted to the IISR inverse problem.

\subsection{Full-Profile Analysis}
In the full-profile analysis method, experimental data $y_{\text{exp}}$ are represented as a correlation profile $\rho_{12}=\langle s(t_1)s(t_2) \rangle$, calculated as the averaged product of the complex quadratures of the received signal. To describe the relationship between the measured signal correlation profiles and the ionospheric plasma parameters (the forward model $y_{\text{model}}$), the radar equation is used \cite{Lehtinen1986, Lehtinen1996}. In the general case, the expected value of the lag product is described via a two-dimensional ambiguity function $W_{t_1,t_2}(r, \tau)$, which defines the spatial-temporal smoothing of responses. However, for the ionic component of the incoherent scatter spectrum, the correlation time of plasma fluctuations significantly exceeds the temporal width of the receiver impulse response. This allows transitioning to the range-reduced ambiguity function $W^{\text{range}}_{t_1,t_2}(r, \tau)$:
\begin{equation}
W^{\text{range}}_{t_1,t_2}(r) = \int_{-\infty}^{\infty} d\tau \int_{-\infty}^{\infty} d\nu W_{t_1}^A(\nu, r) \overline{W_{t_2}^A(\nu-\tau, r)},
\label{AmbiFunc}
\end{equation}
where the amplitude ambiguity function $W_t^A(\tau, r) = h(t-\tau)p(t-2r/c)$ is the product of the receiver filter impulse response $h(t)$ and the emitted pulse envelope $p(t)$, $c$ is the speed of light, $r$ is the slant range, and $\tau = t_2 - t_1$ is the time delay (lag) between samples $t_2$ and $t_1$. Using the range-reduced ambiguity function allows writing the radar equation for the correlation profile in the form (\cite{Hysell2008}):
\begin{equation}
\langle s(t_1)s(t_2) \rangle = A \int_{-\infty}^{\infty} dr P_0(r) W^{\text{range}}_{t_1,t_2}(r) \sigma_{\text{eff}}(t_1-t_2, r, \mathbf{x}(r)).
\label{radeq}
\end{equation}
Here $P_0(r)$ is the single-electron scattering contribution (detailed in Subsection 1.4), $\sigma_{\text{eff}}(t_1-t_2, r, \mathbf{x}(r))$ is the effective plasma scattering cross-section—a two-dimensional function of range and lag that nonlinearly depends on the target plasma parameters $\mathbf{x}(r) = [N_e(r), T_e(r), T_i(r)]^T$ (see Subsection 1.3)—and coefficient $A$ incorporates all radar system constants (transmitter power $P_t$, antenna gain $G_0$, receiver transfer coefficients, and feeder attenuation) that remain constant during a single observation session.

Figure~1 shows a time--range diagram illustrating the physical meaning of ambiguity functions. The main panel depicts the two-dimensional ambiguity function $W_{t,t'}(\tau, r)$, which sets the weighting window for elementary plasma contributions. Parallel to the range axis ($Y$-axis), a typical power profile with characteristic Faraday fading obtained by integration over signal scans is shown in red. Parallel to the time axis ($X$-axis), the emitted pulse envelope $p(t)$ and filter impulse response $h(t)$ are shown in pink, along with the curve $s(t)$ representing raw complex signal quadratures for a single scan, illustrating its stochastic nature before averaging.

Integrating the two-dimensional function along the respective axes yields the range-reduced ambiguity function ($W^{\text{range}}_{t,t'}$) and the lag-reduced ambiguity function ($W^{\text{lag}}_{t,t'}$). As shown in \cite{Lehtinen1986}, simultaneous use of both approximations enables the transition to standard gate analysis. However, this approach is rarely used today because for long single-pulse experiments it fails to correctly account for range smearing. The situation is further complicated by rapid Faraday fading, which renders the inverse problem highly unstable. In this work, we adopt the range-reduced ambiguity function approximation, integrating range smearing effects directly into the forward model \eqref{radeq} within full-profile analysis \cite{Hysell2008, Lehtinen1996}.

According to the Wiener--Khinchin theorem, the effective plasma scattering cross-section $\sigma_{\text{eff}}(\tau, \mathbf{x})$ is related to the spectral scattering density $S(k, \omega)$ via the Fourier transform over time lag $\tau$:
\begin{equation}
\sigma_{\text{eff}}(\tau, r, \mathbf{x}(r)) = \int_{-\infty}^{\infty} d\omega e^{i\omega \tau} S(\omega, r, \mathbf{x}(r)),
\label{Weiner-Hicnhin}
\end{equation}
where the vector $\mathbf{x}$ includes ionospheric plasma parameters: $N_e, T_e, T_i$, etc. The derivation and analysis of the spectral density function are given in Subsection 1.3.

\subsection{Plasma Scattering Cross-Section}
For a collisionless and unmagnetized plasma, the spectral density function takes the form \cite{Froula2011}:
\begin{equation}
S(\omega, r, \mathbf{x}(r)) = \frac{2\pi}{k} |1 - G_e/\epsilon|^{2} f_{e0}(\omega/k) + \frac{2\pi Z}{k} |G_i/\epsilon|^{2} f_{i0}(\omega/k),
\label{Spectrum}
\end{equation}
where $\epsilon = 1 + G_e + G_i$ is the longitudinal dielectric permittivity, and $f_{e0}, f_{i0}$ are Maxwellian particle velocity distributions. The parameters $G_e$ and $G_i$ depend on the ratio of the Debye length $D$ to the wavenumber ($\alpha = 1/kD$) and Faddeeva functions (plasma dispersion function) \cite{Faddeeva1954, Fried1961} $W(\chi) = R_w(\chi) + i I_w(\chi)$:
\begin{equation}
D = \sqrt{\frac{\epsilon_0 \kappa T_e}{e^2 N_e}}, \quad R_w(\chi) = 1 - 2\chi e^{-\chi^2} \int_{0}^{\chi} e^{p^2} dp, \quad I_w(\chi) = \sqrt{\pi} \chi e^{-\chi^2}.
\end{equation}
The complex susceptibility parameters of electron ($G_e$) and ion ($G_i$) plasma components are:
\begin{equation}
G_e = \alpha^2 W(\chi_e), \quad G_i = \alpha^2 Z \frac{T_e}{T_i} W(\chi_i).
\end{equation}
Here $\chi_e = \omega/ka$ and $\chi_i = \omega/kb$ are normalized frequencies, where $a$ and $b$ are thermal particle velocities ($\epsilon_0$ is the vacuum permittivity, $\kappa$ is Boltzmann's constant, $e$ is electron charge), and $Z$ is ion charge ($Z = 1$ for singly ionized $\text{O}^+$ ions). In the Salpeter approximation \cite{Froula2011, Salpeter1960}, the ionic component of the spectrum (ion line) simplifies to:
\begin{equation}
S_i(k, \omega) \equiv \frac{2\pi^{1/2}}{kb} \exp(-\chi_i^2) N_e \left|1 + \alpha^{-2} + \frac{T_e}{T_i} W(\chi_i)\right|^{-2}.
\label{Spectrum_Salpeter}
\end{equation}

To optimize the solution of the inverse problem, the scattering cross-section is factorized into two independent terms \cite{Dougherty1960}:
\begin{equation}
\sigma_{\text{eff}}(\tau, r, \mathbf{x}(r)) = \sigma_{\text{eff}}(0, r, N_e(r)) \cdot \tilde{\sigma}_{\text{eff}}(\tau, r, T_e(r), T_i(r)).
\label{eq:sigma}
\end{equation}
The first factor $\sigma_{\text{eff}}(0, r, N_e(r))$ corresponds to zero lag ($\tau=0$) and depends only on electron density $N_e(r)$ (assuming a constant $T_e(r)/T_i(r)$ ratio), while $\tilde{\sigma}_{\text{eff}}(\tau, r, T_e(r), T_i(r))$ is the normalized scattering cross-section function whose shape is determined exclusively by plasma temperatures.

\subsection{Wideband Signal Analysis}
The total scattering cross-section at zero lag $\sigma_{\text{eff}}(0, r, \mathbf{x}(r))$ is defined by the expression:
\begin{equation}
\sigma_{\text{eff}}(0, r, \mathbf{x}(r)) = \frac{N_e(r)}{(1+(kD(r))^2)(1+(kD(r))^2 + T_e(r)/T_i(r))},
\end{equation}
which for $kD(r) \ll 1$ (which is valid for ionospheric conditions and radar wavelengths) can be simplified as:
\begin{equation}
\sigma_{\text{eff}}(0, r, \mathbf{x}(r)) = \frac{N_e(r)}{1 + T_e(r)/T_i(r)},
\end{equation}
where $D(r) = \sqrt{\epsilon_0 \kappa T_e(r) / e^2 N_e(r)}$ is the Debye length. The dependence in the numerator reflects the power growth proportional to the particle number, while the denominator accounts for Debye shielding effects and the influence of the temperature ratio $T_e/T_i$, through which ions control electron motion. The wideband signal is used to measure the scattering power with high spatial resolution. The measured power is related to the total scattering cross-section via the radar equation \eqref{radeq}, which, under the assumption $T_e/T_i = 1$, allows reducing the inverse problem to recovering the electron density profile $N_e(r)$.

For a wideband channel with Barker coding, the spatial resolution is determined by the short duration of the code sub-pulse. This allows neglecting the range extent of the ambiguity function $W_r(r)$ and approximating it with a Dirac delta function $\delta(r - r')$ \cite{Alsatkin2020}. Such an approximation reduces the integral radar equation \eqref{radeq} to a local power model $p_{\text{exp}}(t) = \langle s^2(t) \rangle$ (see Fig.~\ref{fig:power}):
\begin{equation}
p_{\text{model}}(r) = A \cdot P_0(r) \cdot \sigma_{\text{eff}}(0, x(r), r) + B,
\label{eq:wb_power_model}
\end{equation}
where $\sigma_{\text{eff}}(0, r, \mathbf{x}(r))$ is the effective plasma scattering cross-section at $\tau=0$, $A$ is the radar scaling factor, and $B$ is the additive sky noise contribution, which are calculated directly as linear regression coefficients, while $P_0(r)$ is the single-electron scattering contribution for a monostatic radar with a single linear polarization:
\begin{equation}
P_0(r) = \frac{r_0^2 c^2 \lambda^2 \cos^2 \Omega(r) P_t G_0^2(r)}{(16\pi r^2)^2}.
\label{eq:P0}
\end{equation}
Here $r_0$ is the classical electron radius, $c$ is the speed of light, $\lambda$ is the transmitter wavelength, $P_t$ is the radiated power, $G_0(r)$ is the antenna gain, and the factor $\cos^2 \Omega(r)$ describes power fading due to Faraday polarization plane rotation \cite{Shpynev2004, Shpynev2017}. The rotation angle $\Omega(r)$ is determined by the integrated electron density along the ray path:
\begin{equation}
\Omega(r) = \gamma \int_0^r N_e(z) \, dz, \quad \gamma = \frac{e^3 B_0 \cos \alpha}{2 \epsilon_0 m_e^2 \omega_0^2 c},
\label{eq:faraday_angle}
\end{equation}
where $B_0$ is the geomagnetic field induction, $\alpha$ is the angle between the wave vector and the magnetic field, and $\omega_0$ is the radar carrier frequency.

The electron density profile model $N_e(r)$ is specified by a Chapman $\beta$-layer:
\begin{equation}
N_e(r; N_m, h_m, H_b, H_t) = N_m \exp\left( 1 - z - e^{-z} \right), \quad z = \frac{r - h_m}{H}, H(r) = \begin{cases} H_b; r \le h_m, \\ H_t; r &gt; h_m, \end{cases}
\label{eq:chapman_layer}
\end{equation}
where $N_m$ is the peak concentration, $h_m$ is the layer peak height, and $H_b, H_t$ are the scale heights of the homogeneous atmosphere.

The non-linear optimization problem consists in fitting the experimental power profile $p_{\text{exp}}(r)$ by searching for the Chapman layer parameters $(N_m, h_m, H_b, H_t)$ while imposing empirical two-sided bounds on the coefficient $A$:
\begin{equation}
A_{\min} \le A \le A_{\max}.
\label{eq:box_constraints_A}
\end{equation}
This problem is solved using a global optimization algorithm \cite{Tashlykov2024} or by grid search over the parameter space \cite{Alsatkin2020}. The output of this stage is the primary profile $P_0(r)$, which is required to initialize the narrowband signal analysis. In addition, the time series of the coefficient $A$ prove to be extremely important, as their analysis provides supplementary information on the dynamics of the temperature ratio profile $T_e/T_i$.

Let us return to the integral signal power equation and rewrite it as:
\begin{equation}
p(r) = A' \int_{r_{\min}}^{r_{\max}} W_r(r') \frac{\cos^2 \Omega(r')}{(r')^2} N_e(r') \frac{1}{1 + T_e(r')/T_i(r')} \, dr'.
\label{eq:integral_power_raw}
\end{equation}
Since the factor $\frac{1}{1 + T_e(r')/T_i(r')}$ is continuous and strictly positive over the entire integration interval $[r_{\min}, r_{\max}]$, the first mean value theorem for definite integrals can be applied to \eqref{eq:integral_power_raw}. As a result, we obtain:
\begin{equation}
p(r) = A' \cdot \frac{1}{1 + T_e(\xi)/T_i(\xi)} \int_{r_{\min}}^{r_{\max}} W_r(r') \frac{\cos^2 \Omega(r')}{(r')^2} N_e(r') \, dr',
\label{eq:mean_value_applied}
\end{equation}
where $\frac{1}{1 + T_e(\xi)/T_i(\xi)}$ is the value of the temperature factor at some intermediate point $\xi \in [r_{\min}, r_{\max}]$, and the factor $A'$ contains all radar constants.

Once the optimal profile $N_e(r)$ is recovered, the value of $P_0(r)$ is calculated directly and used as an input parameter in the second inverse problem—recovering plasma temperatures in the narrowband channel. The scaling coefficient $A$, connecting the experimental and model power, explicitly depends on the temperature ratio at a fixed height $r_\zeta$: $A = A' \frac{2}{1 + T_e(r_\zeta)/T_i(r_\zeta)}$. Provided that the radar system constants $A'$ remain unchanged during a single continuous observation session, this allows evaluating a reference baseline value $T_e/T_i$, which is implemented as an additional non-linear constraint $\frac{T_e(r_{\max})}{T_i(r_{\max})} \ge \frac{T_e(r_\xi)}{T_i(r_\xi)}$ for the optimization problem at the narrowband signal analysis stage. This addition significantly stabilizes the solution search and eliminates numerically feasible but physically inadequate results.

\subsection{A Priori Constraints and Regularization Strategy}
To ensure the physical adequacy of the solution, the problem is supplemented by a system of a priori constraints. The temperature profiles are approximated by a linear combination of 3rd-order de Boor basis splines $B_j^3(r)$, where the target optimization parameters are the coefficients (weights) $a_j$ (for $T_e$) and $b_j$ (for $T_i$):
\begin{equation}
T_e(r) = \sum_{j=1}^n a_j B_j^3(r), \quad T_i(r) = \sum_{j=1}^n b_j B_j^3(r).
\label{eq:Bsplines}
\end{equation}

The constraint system includes the following conditions:
\begin{itemize}
\item Search range: Coefficient values are bounded by physically probable limits $300 \le a_j, b_j \le 4000$~K.
\item Thermal equilibrium at boundaries: At the lower boundary (100~km), $T_e \approx T_i$ is assumed due to high collision frequencies in the D-region. At the upper boundary (1000~km), $T_e \approx T_i$ is also adopted as an a priori trend consistent with satellite, rocket, and radar observations for the quiet ionosphere.
\item Physical inequalities: Conditions $T_e(r) \ge T_i(r)$, or $a_j \ge b_j$, are imposed for all altitudes, along with a non-decreasing condition for ion temperature with altitude ($b_j \le b_{j+1}$).
\item Wideband channel stabilization: An additional non-linear constraint is used:
\begin{equation}
\frac{T_e(r_{\max})}{T_i(r_{\max})} \ge \frac{T_e(r_\xi)}{T_i(r_\xi)},
\label{eq:anchor_inequality}
\end{equation}
where the reference value on the right-hand side is evaluated from the scaling coefficient $A$ obtained from the preliminary $N_e$ analysis.
\item To ensure smoothness and controlled curvature of the temperature ratio profile $T_e/T_i$ (based on Jensen's inequality), a discrete condition is imposed on the ratio differences of the B-spline coefficients for internal nodes ($j = 2, \dots, n-1$):
\begin{equation}
2 \frac{a_j}{b_j} - \frac{a_{j+1}}{b_{j+1}} - \frac{a_{j-1}}{b_{j-1}} \le 0.
\label{eq:discrete_curvature}
\end{equation}
Relation \eqref{eq:discrete_curvature} regulates the local convexity of the sequence of reference values $u_j = a_j/b_j$, effectively suppressing oscillations and eliminating unphysical temperature ratio outliers at individual altitude levels.
\end{itemize}

\subsection{Global Optimization Algorithm and Regularization Strategy}
To solve the minimization problem of the objective functional \eqref{eq:objective_func}, the ISRES global search and COBYLA local optimization algorithms implemented in the NLopt library \cite{NLopt} were used. The cost surface of the residual functional $\Psi(\mathbf{x})$ in the temperature recovery problem is characterized by clusters of local extrema and stretched global minimum regions \cite{Shpynev2010, Vallinkoski1988}. To solve this ill-posed problem, the Improved Stochastic Ranking Evolution Strategy (ISRES) \cite{Runarsson2000, Runarsson2005} is applied. The algorithm operates on a population of "parent" solutions, generating "offspring" through crossover and mutation mechanisms, which provides a reliable escape from local minima without requiring a precise initial guess.

An empirically established stopping criterion is completing 10,000 iterations, which guarantees reaching the global optimum and accelerates calculations by four orders of magnitude relative to the grid search method. Non-linear constraints are handled via the stochastic ranking strategy \cite{Runarsson2000, Runarsson2005}, whereas local solution refinement is performed using the COBYLA linear approximation method \cite{Powell1994, Powell1998}.

To ensure the physical smoothness of time series and eliminate unphysical $T_i(r, t)$ oscillations, the entire optimization process is divided into three sequential stages:
\begin{enumerate}
\item Preliminary stage (Wideband channel): Solving the inverse problem for the Barker-coded pulse allows recovering the $N_e(r)$ profile, calculating the $P_0(r)$ contribution, and determining the reference ratio $T_e(r_\xi)/T_i(r_{\xi})$ by analyzing the time series of the scaling coefficient $A$ (under the assumption that the maximum value of $A$ during continuous measurements corresponds to the nighttime condition $T_e/T_i=1$).
\item Global search (Narrowband channel): Searching for the global minimum of $\Psi$ using the ISRES algorithm (10,000 iterations). At this stage, the general structure of the $T_e$ and $T_i$ profiles is formed, incorporating the non-linear temperature ratio constraint derived from wideband analysis.
\item Temporal regularization: Refining the $T_i$ profile using the local COBYLA algorithm with a variability constraint (no more than 20\% from the best solution over a 30-minute window), which ensures the physical smoothness of the time series.
\end{enumerate}

The two-step "ISRES $\to$ COBYLA" scheme yields relatively smooth, physically adequate altitude-temporal temperature distributions while increasing the overall computational complexity of the algorithm by no more than 20\%.

\subsection{Parameter Error Estimation}
To assess the accuracy of recovering temperature profiles, a local linearization method of the forward problem model near the found global minimum is used \cite{Hysell2008, Vierinen2012}. The a posteriori covariance matrix of the parameters $\Sigma_x$ is calculated based on the Jacobian matrix $J$ and the data covariance matrix $\Sigma_d$ \cite{Bevington2003}:
\begin{equation}
\Sigma_x = (J^T \Sigma_d^{-1} J)^{-1},
\label{eq:cov_param}
\end{equation}
where the elements of $J$ represent partial derivatives of the model with respect to the spline weights \cite{Hysell2008, Holt1992}. The elements of the matrix $\Sigma_d$ between two lag products $\rho_{12}$ and $\rho_{34}$ are determined based on the fourth moments of complex Gaussian quantities \cite{Hysell2008}:
\begin{equation}
C_d(12,34) = \frac{1}{2k} \mathrm{Re}(\langle\rho_{13}\rangle\langle\rho_{24}^*\rangle + \langle\rho_{23}\rangle\langle\rho_{14}^*\rangle),
\end{equation}
where $k$ is the number of independent scans \cite{Hysell2008, Lehtinen1986}.

The computational efficiency of the algorithm is achieved by utilizing the local support property of B-splines \cite{Holt1992, Lehtinen1986}. Since each basis spline is non-zero only over a bounded altitude interval, the Fisher information matrix $Q = \Sigma_x^{-1}$ is formed by block-wise accumulation of contributions from individual altitude intervals:
\begin{equation}
Q = \sum_{a} J_{a}^T \Sigma_{d,a}^{-1} J_{a},
\end{equation}
where $J_a$ and $\Sigma_{d,a}$ are the local Jacobian and covariance for a specific altitude $a$. This approach avoids the inversion of ultra-large data matrices ($\sim 3 \cdot 10^5$ points), reducing the problem to working with small blocks while preserving all statistical information \cite{Hysell2008, Vierinen2012}.

The relationship between the measurement structure and their errors is illustrated in Fig.~\ref{fig:covariance}. The left panel shows an example of the real part of an experimental correlation profile in the "altitude--lag" axes. The right panel of Fig.~\ref{fig:covariance} presents a block of the data covariance matrix for an altitude of 250~km. Owing to the high temporal resolution ($\Delta t = 4~\mu\text{s}$), the correlation structure within the pulse appears as quasi-continuous diagonal bands, in contrast to the discrete "cells" characteristic of experiments with a coarse sampling step \cite{Hysell2008}. Despite the completeness of the statistical accounting, such a linearized estimate may remain "optimistic" due to the strong mutual correlation between $T_e$, $T_i$, and the ion composition, which requires caution when interpreting the obtained confidence intervals \cite{Vallinkoski1988, Andrae2010}.

\subsection{Physical Approximations and Error Estimation}

In formulating the forward problem, two key physical assumptions were adopted. First, the plasma is assumed to be single-component, consisting exclusively of $\text{O}^+$ ions. Introducing ion composition as an additional target variable under IISR conditions leads to further destabilization of the inverse problem due to the lack of unambiguous signatures of ion composition in correlation profiles and the risk of the TICA (temperature-ion composition ambiguity) effect \cite{Martinez-Ledesma2019}. We intentionally refrain from using external models, such as IRI \cite{Bilitza2017}, to fix the composition, striving for the "self-sufficiency" of the IISR diagnostic technique. Using full-profile analysis (FPA) \cite{Holt1992, Lehtinen1986, Hysell2008} allows partially compensating for this simplification by accounting for vertical parameter correlations. Errors associated with the influence of molecular ions in the lower F-region will be considered in detail in future works. The strong error correlation of IS parameters limiting the accuracy of multi-parameter fitting was theoretically justified in \cite{Vallinkoski1988}, while the impact of random noise on drift and temperature estimates was addressed in \cite{Williams1996}.

Second, this model does not account for the Doppler shift of the spectrum caused by plasma drift. Traditionally, determining the drift velocity requires analyzing the imaginary part of the correlation function, which is characterized by a low signal-to-noise ratio (SNR) and requires significant accumulation time, during which plasma parameters can change substantially \cite{Shcherbakov2009}. To justify the applicability of this assumption, a numerical sensitivity analysis was conducted. A Doppler shift $\omega \to \omega - k V_d$ was introduced into the theoretical formula for the ion line spectral density (equation \eqref{Spectrum_Salpeter}). For each value of velocity $V_d$ in steps of 10~m/s, a pair $\{T_e, T_i\}$ was found using the "nearest neighbor" algorithm (with a temperature step of 25~K) that best described the shifted spectrum within the zero-drift model.

The results of the temperature recovery error estimation are presented in Figures~\ref{fig:Err_image} and \ref{fig:Err_image_Te}:

\begin{itemize}
\item Figure~\ref{fig:Err_image}: Dependence of the $T_e$ and $T_i$ recovery error on the drift velocity $V_d$ and ion temperature $T_i$ for a fixed electron temperature $T_e = 3000$~K. The figure demonstrates that at typical mid-latitude velocities ($V_d < 100$~m/s), the error remains within 5\%, whereas at extreme drift values, the $T_i$ error can reach significant levels.
\item Figure~\ref{fig:Err_image_Te}: Dependence of the $T_e$ and $T_i$ recovery error on the drift velocity $V_d$ and electron temperature $T_e$ for a fixed ion temperature $T_i = 1000$~K. This visualization confirms that the "steepening" of spectral peaks as the $T_e/T_i$ ratio increases affects the stability of parameter recovery, as noted in studies on random noise estimation \cite{Williams1996}.
\end{itemize}

The sensitivity analysis showed that for drift velocities up to 100~m/s (typical for mid-latitudes), the recovery error for $T_e$ and $T_i$ does not exceed 5\%. A significant drift impact ($T_i$ error up to 30\% at $V_d = 300$~m/s) manifests only during periods of extreme geomagnetic activity, which justifies neglecting the drift velocity for automated monitoring of the quiet ionosphere.

\subsection{Unified Mathematical Model}
To eliminate redundant formula duplication, the final problem of recovering ionospheric temperatures $T_e(r)$ and $T_i(r)$ is formulated as a unified non-linear mathematical programming system.

The target parameter vector is the vector of B-spline coefficients $\mathbf{x} = [a_1, \dots, a_n, b_1, \dots, b_n]^T \in \mathbb{R}^{2n}$, specifying the altitude temperature profiles \eqref{eq:Bsplines}.

The unified mathematical formulation of the inverse problem synthesizes the forward model of the narrowband channel and penalty functions into the following optimization system:
\begin{equation}
\begin{cases}
\mathbf{x}^* = \arg\min_{\mathbf{x} \in \mathbb{R}^{2n}} \left[ f(\mathbf{x}) + \lambda \sum_{j=1}^m \max\left(0, g_j(\mathbf{x})\right)^2 \right], \\
f(\mathbf{x}) = \sum_{t_1, t_2} \bigg( R_{\text{exp}}(t_1, t_2) - \int_{r_{\min}}^{r_{\max}} \tilde{P}_0(r) W_{t_1,t_2}(r) \tilde{\sigma}_{\text{eff}}(t_1 - t_2; T_e(r;\mathbf{x}), T_i(r;\mathbf{x})) \, dr \bigg)^2, \\
\tilde{P}_0(r) = A' \dfrac{\cos^2\Omega(r)}{r^2} N_e(r),
\end{cases}
\label{eq:unified_system_main}
\end{equation}
subject to a unified system of linear and non-linear constraints $g_j(\mathbf{x}) \le 0$:
\begin{equation}
\begin{cases}
300 - a_j \le 0, \quad a_j - 4000 \le 0; \quad j = 1, \dots, n, \\
300 - b_j \le 0, \quad b_j - 4000 \le 0; \quad j = 1, \dots, n, \\
b_j - a_j \le 0; \quad j = 1, \dots, n, \\
b_j - b_{j+1} \le 0; \quad j = 1, \dots, n-1, \\
a_1 - 1.1 b_1 \le 0; \quad (r = 100~\text{km}), \\
a_n - 1.1 b_n \le 0; \quad (r = 1000~\text{km}), \\
\frac{T_e(r_{\max})}{T_i(r_{\max})} \ge \frac{T_e(r_\xi)}{T_i(r_\xi)}, \\
2 \frac{a_j}{b_j} - \frac{a_{j+1}}{b_{j+1}} - \frac{a_{j-1}}{b_{j-1}} \le 0; \quad j = 2,\dots,n-1.
\end{cases}
\label{eq:unified_constraints_system}
\end{equation}

The interaction logic between channels can be represented as a two-stage algorithm (see Fig.~\ref{fig:flowchart}): first, an independent problem is solved for the wideband signal ($N_e$, $P_0(r)$, baseline $T_e/T_i$), the results of which are passed as constants and constraints to the main global temperature optimization problem in the narrowband channel. In system \eqref{eq:unified_system_main}--\eqref{eq:unified_constraints_system}, the parameters $N_e(r)$, $\Omega(r)$, and the baseline value $\langle T_e/T_i \rangle_{\xi}$ are fixed from the preliminary wideband signal analysis. Thus, the formulated optimization problem is sequentially implemented within a three-stage calculation scheme (see Section~1.6): from preliminary parameter fixing of the wideband channel to stochastic global solution search and its subsequent local regularization.

\section{Results and Discussion}
The initial stage of data processing is the analysis of wideband channel data. Figure~\ref{fig:Tr} presents the diurnal variations of the ionospheric peak parameters ($N_mF_2$, $h_mF_2$) and the time series of the amplitude scaling factor $A$. An automated procedure for estimating $N_e$ based on parametric Chapman layer models and the effective pulse subtraction method was previously successfully tested at the IISR \cite{Alsatkin2020, Tashlykov2024}. Using the generalized mean value theorem for the radar equation, we obtain the temperature ratio value at a reference altitude $1/(1+T_e(r_\xi)/T_i(r_\xi))$. As shown in the plot, this parameter serves as a stable "anchor" that bounds the search space in the second stage. It is important to note that although the exact altitude $r_\xi$ from the generalized mean value theorem is not specified, the value $T_e(r_\xi)/T_i(r_\xi)$ is guaranteed to lie below the maximum of the temperature ratio profile, making it a reliable "anchor" for narrowing the search domain. Without applying this constraint, we observe systematically overestimated daytime $T_i$ values, leading to an unnaturally sharp contrast between daytime and nighttime temperatures. This is a long-standing issue in IISR data caused by the "stretched" topology of the residual functional global minimum \cite{Shpynev2010}. Unlike the $N_e$ analysis, where errors are visible through shifts in Faraday fading nodes, in the narrowband channel such shifts lack an explicit visual signature. This confirms that the stability of the solution in the "off" mode is a numerical artifact of the "stretched" minimum, which is overcome only by introducing equation \eqref{eq:anchor_inequality}. In the "on" mode, daytime $T_i$ values become noticeably lower, and the diurnal variation becomes more physically reasonable. The use of B-splines guarantees profile smoothness, while the "anchor" from the wideband channel effectively eliminates false minima into which unconstrained search traps.

Temperature dynamics at altitudes of 250, 300, 350, and 400~km for the period July 10--13 are presented in Figures~\ref{fig:res_250}, \ref{fig:res_300}, \ref{fig:res_350}, and \ref{fig:res_400}, comparing the values obtained in the "on" and "off" modes. With the constraint enabled ("on" mode), daytime $T_i$ values become substantially lower (e.g., by 50\% on average at an altitude of 300~km). This resolves previously observed discrepancies between IISR data and measurements from other mid-latitude radars \cite{Goncharenko2005}. The difference between daytime maximum and nighttime minimum $T_i$ decreases to values consistent with modern ionospheric models \cite{Bilitza2017, Zhang2005, Zhang2007}. To obtain smooth time series, a two-step scheme is used: the ISRES algorithm establishes the profile structure, after which the COBYLA algorithm acts as a temporal regularizer. The search domain for COBYLA is constrained to a 20\% corridor from the best result over a 30-minute interval, which effectively suppresses noise while preserving real ionospheric dynamics.

The resulting 2D maps of signal-to-noise ratio, $T_e$, and $T_i$ exhibit smoothness and an absence of anomalous gradients (see Fig.~\ref{fig:res_image}). Sensitivity analysis (Figs.~\ref{fig:Err_image} and \ref{fig:Err_image_Te}) confirms that at typical velocities $V_d < 100$~m/s, the temperature recovery error does not exceed 5\%, which is comparable to the instrumental noise level. It should be emphasized that this paper focuses on describing the automated methodology; detailed comparison of the recovered data with the IRI-2020, ISRIM, and PIM models \cite{Bilitza2017, Zhang2005, Zhang2007, Romanova2008} and with independent measurements of neutral temperature ($T_n$) by a Fabry--Perot interferometer (FPI) \cite{Vasilyev2017} will be presented in a subsequent publication.

\section{Conclusion}
This paper presents a new automated method for recovering ionospheric plasma temperature profiles from Irkutsk Incoherent Scatter Radar (IISR) data. The core achievement is the development of a two-stage computational algorithm that integrates data from wideband and narrowband channels to compensate for distortions caused by Faraday rotation on a radar with a single linear polarization antenna. Key findings of the study:
\begin{itemize}
\item Solution Stabilization: Utilizing wideband channel data (Barker code) to establish an "anchor" constraint on the $T_e/T_i$ ratio effectively resolves the issue of the "stretched" global minimum of the objective functional. Experiments demonstrate that without this constraint, the algorithm systematically converges to stable yet physically overestimated daytime ion temperature values.
\item Optimization Efficiency: Combining global stochastic search (ISRES) and local temporal regularization (COBYLA) provides reliable recovery of smooth parameter profiles without requiring manual initial guesses.
\item Error Assessment: Numerical sensitivity analysis confirms that for the quiet mid-latitude ionosphere, neglecting plasma drift ($V_d < 100$~m/s) and ion composition is a justified simplification, as the resulting systematic error does not exceed 5\%, which is comparable to instrumental noise levels.
\end{itemize}
The developed method establishes IISR measurements as a self-sufficient tool for long-term ionospheric monitoring. Future work will focus on incorporating ion composition into the set of fitted variables and performing detailed data validation through comparison with empirical ionospheric models \cite{Bilitza2017, Zhang2005, Romanova2008, Vasilyev2025} and independent neutral temperature measurements by a Fabry--Perot interferometer \cite{Vasilyev2017}.

\section{Acknowledgments}
This work was supported by the Ministry of Science and Higher Education of the Russian Federation. Results were obtained using the Unique Scientific Facility "Irkutsk Incoherent Scatter Radar" of ISTP SB RAS (https://ckp-rf.ru/catalog/usu/77733/).

\newpage
\begin{figure}[h!]
\centering
\includegraphics[width=0.8\textwidth]{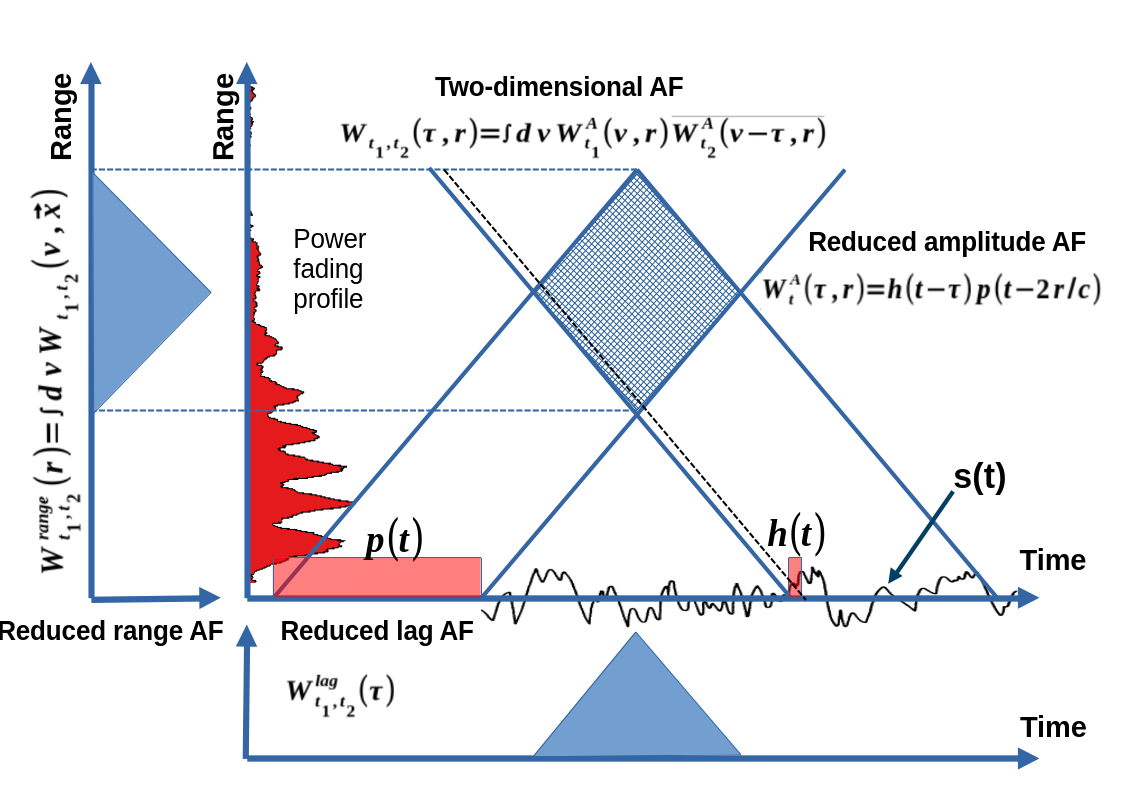}
\caption{Time--range diagram ($t$--$S$) for a single-pulse experiment. For the IISR, the characteristic transmitted pulse length $p(t)$ is 700~$\mu\text{s}$, and the impulse response duration $h(t)$ is 4~$\mu\text{s}$.}
\label{fig:ambiguity_diagram}
\end{figure}

\newpage
\begin{figure}[h!]
\centering
\includegraphics[width=0.8\textwidth]{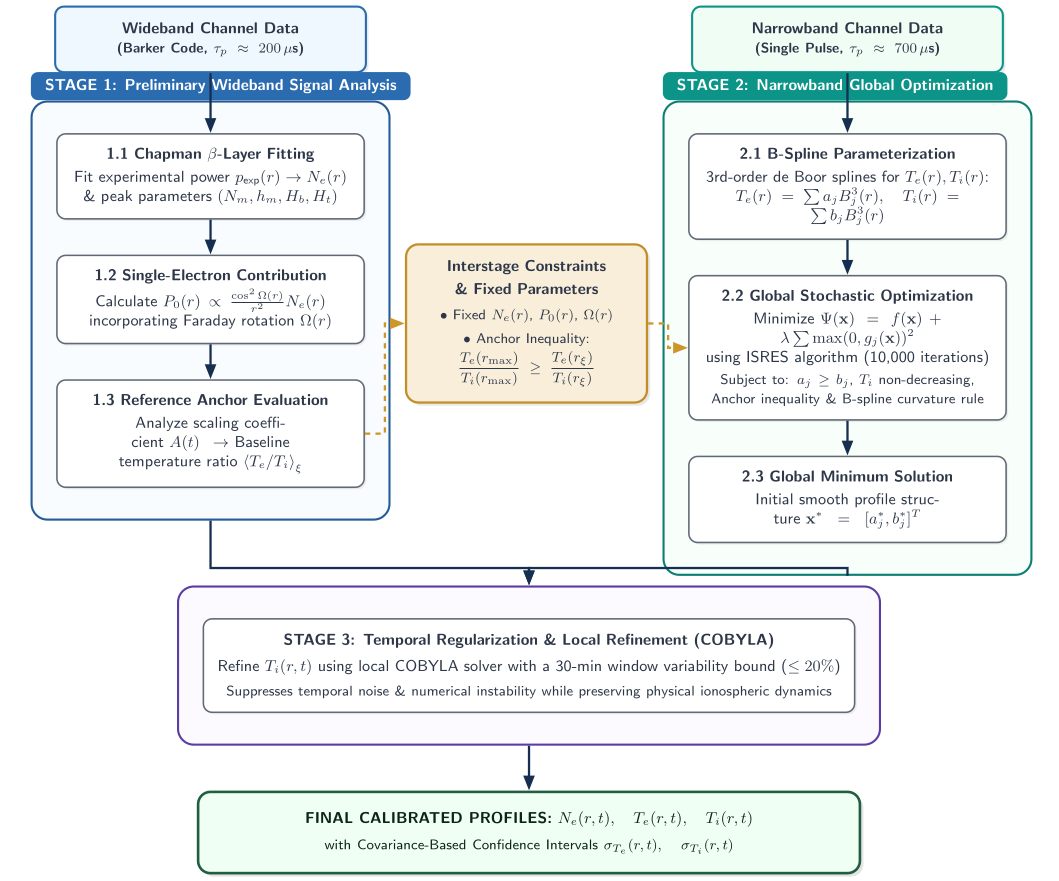}
\caption{Flowchart of the temperature recovery algorithm: from preliminary wideband signal analysis to two-stage profile regularization.}
\label{fig:flowchart}
\end{figure}

\newpage
\begin{figure}[h!]
\centering
\includegraphics[width=0.8\textwidth]{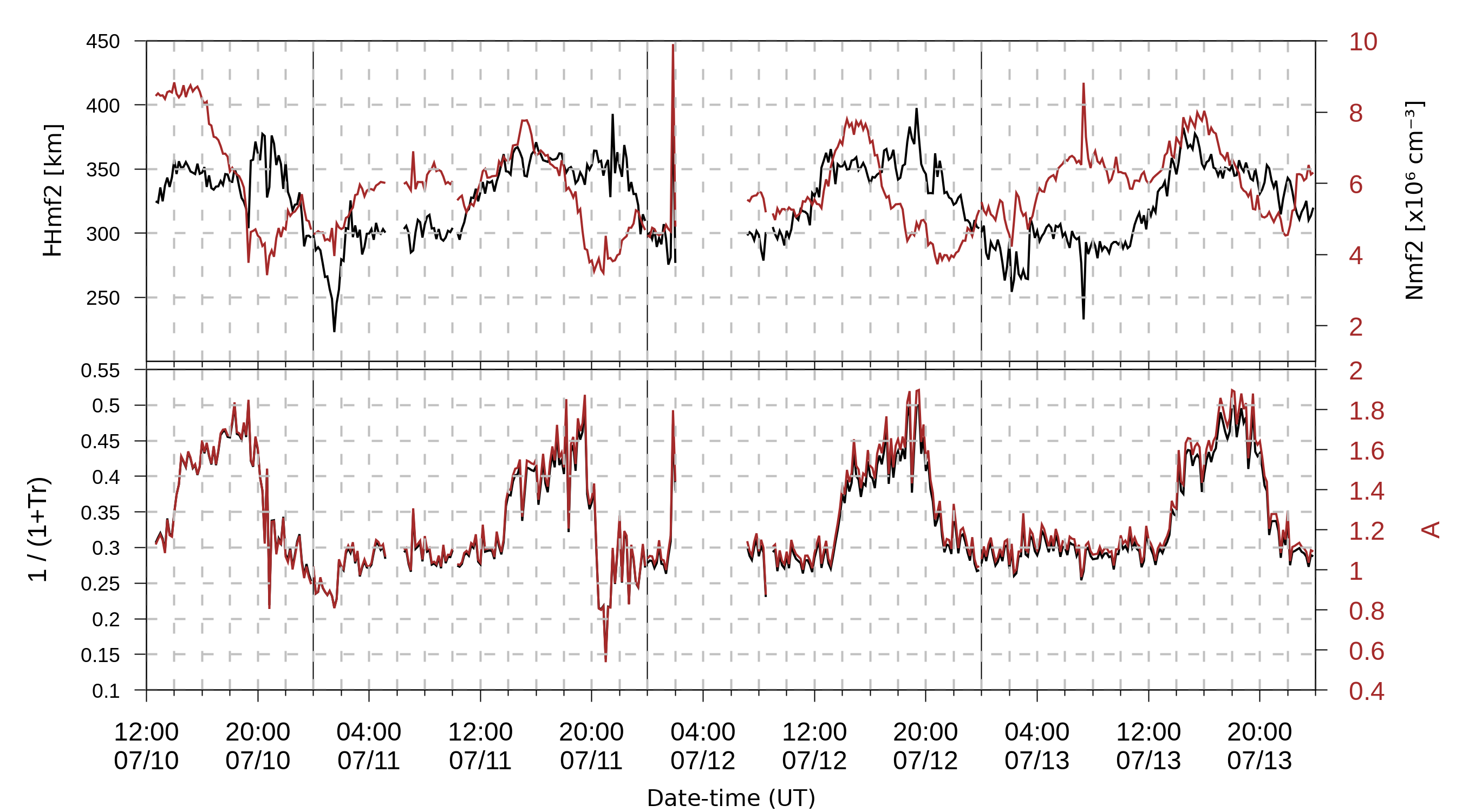}
\caption{Results of wideband signal analysis (Channel 1, frequency 159.8~MHz, accumulation time 10 minutes). Top panel: comparison of key $N_e$ profile peak parameters (peak altitude $h_mF_2$ and peak concentration $N_mF_2$). Bottom panel: baseline value $1/(1+T_e(r_\xi)/T_i(r_\xi))$ obtained by analyzing scaling factor $A$ under the assumption that the minimum value of the diurnal $T_e(r)/T_i(r)$ variation equals 1, while other radar parameters (see equation \eqref{eq:P0}) remain constant.}
\label{fig:Tr}
\end{figure}

\newpage
\begin{figure}[h!]
\centering
\includegraphics[width=0.8\textwidth]{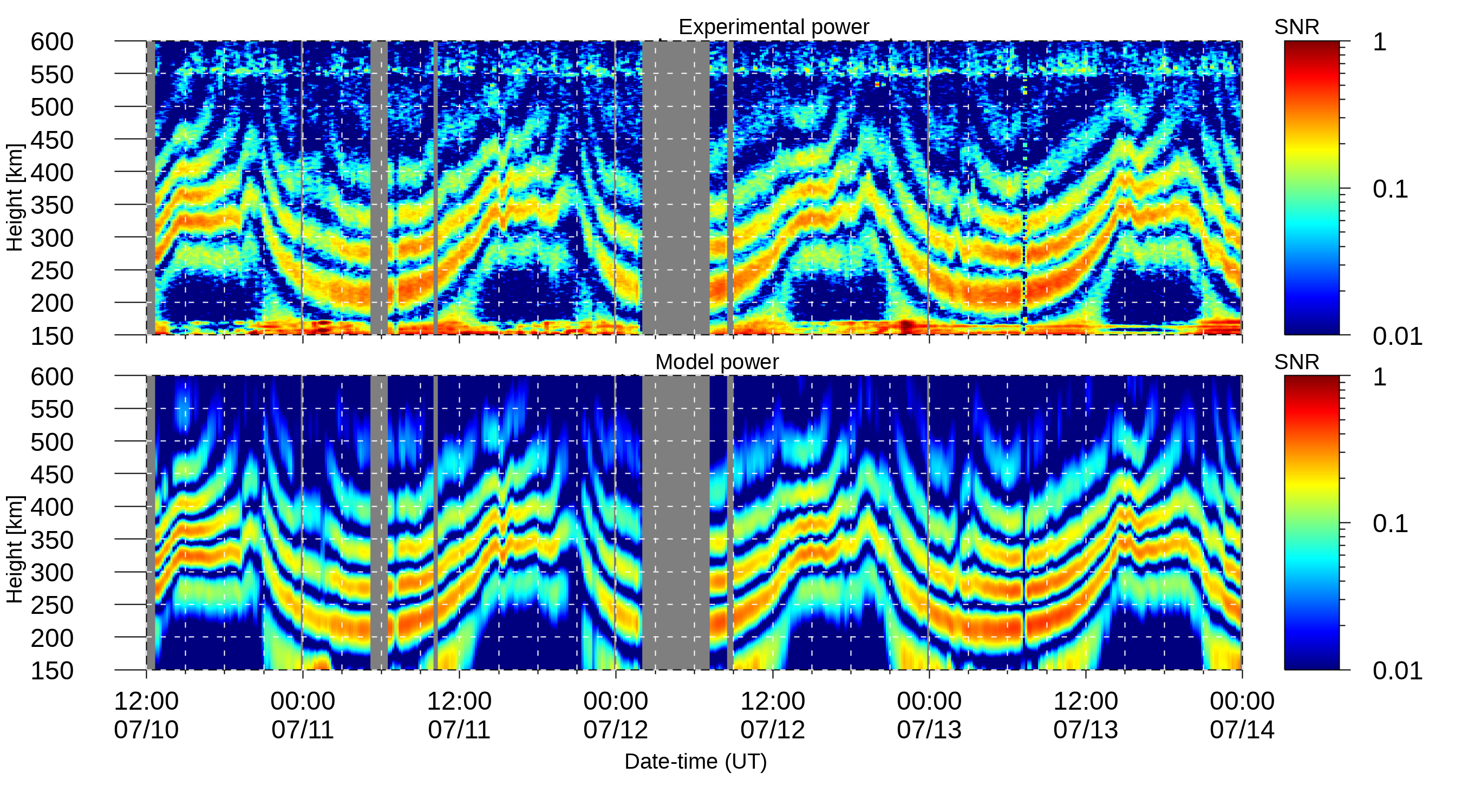}
\caption{Comparison of experimentally measured power (top panel: Channel 1, frequency 159.8~MHz, accumulation time 10 minutes) and model power obtained via nonlinear optimization (bottom panel).}
\label{fig:power}
\end{figure}

\newpage
\begin{figure}[h!]
\centering
\begin{minipage}{0.8\textwidth}
\includegraphics[width=\textwidth]{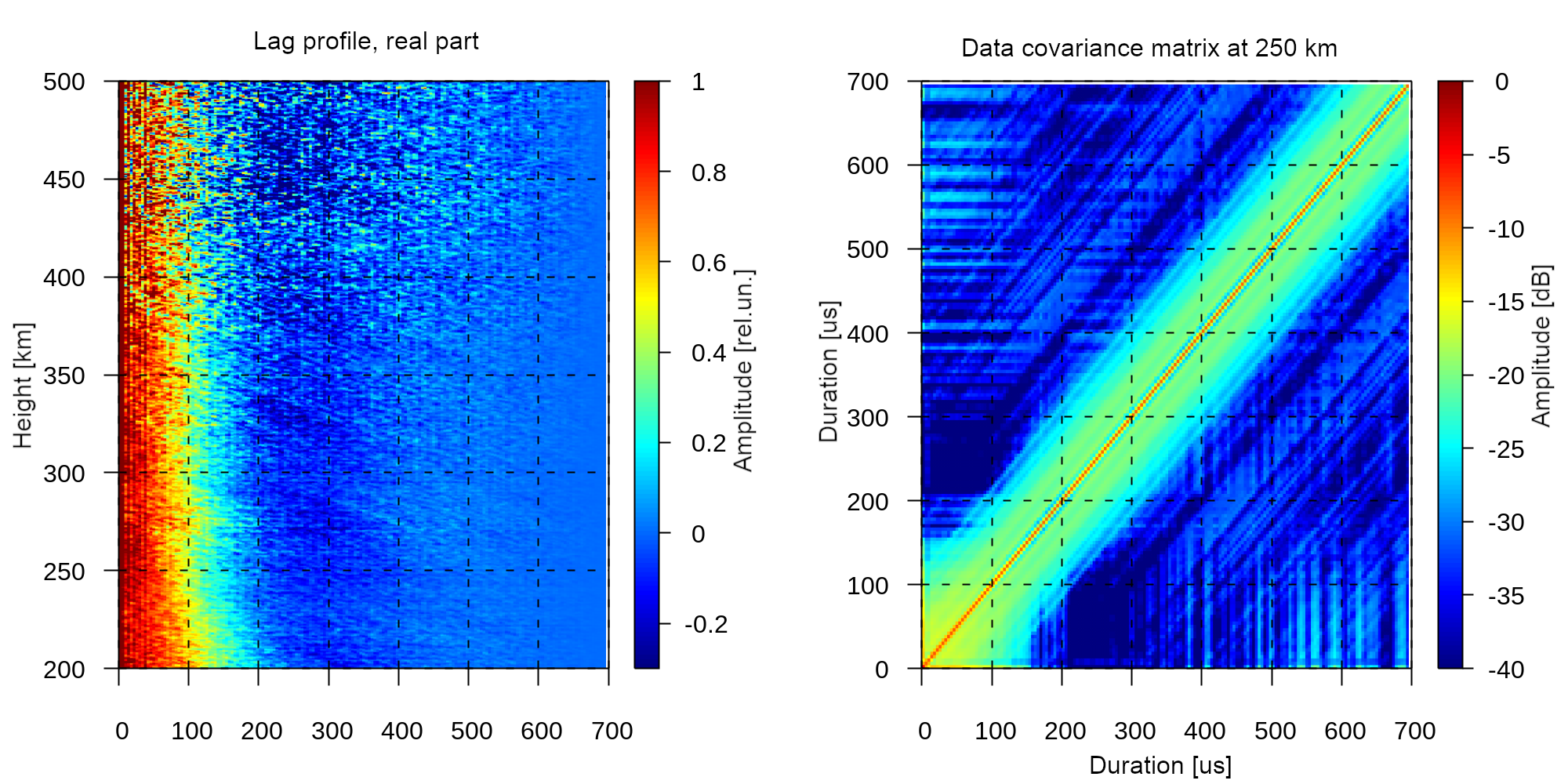}
\end{minipage}
\caption{Left: example of the real part of an experimental correlation profile for 2023-07-13 05:00 UT measurements (2000-scan accumulation). Right: block of the data covariance matrix for an altitude of 250~km.}
\label{fig:covariance}
\end{figure}

\newpage
\begin{figure}[h!]
\centering
\includegraphics[width=0.8\textwidth]{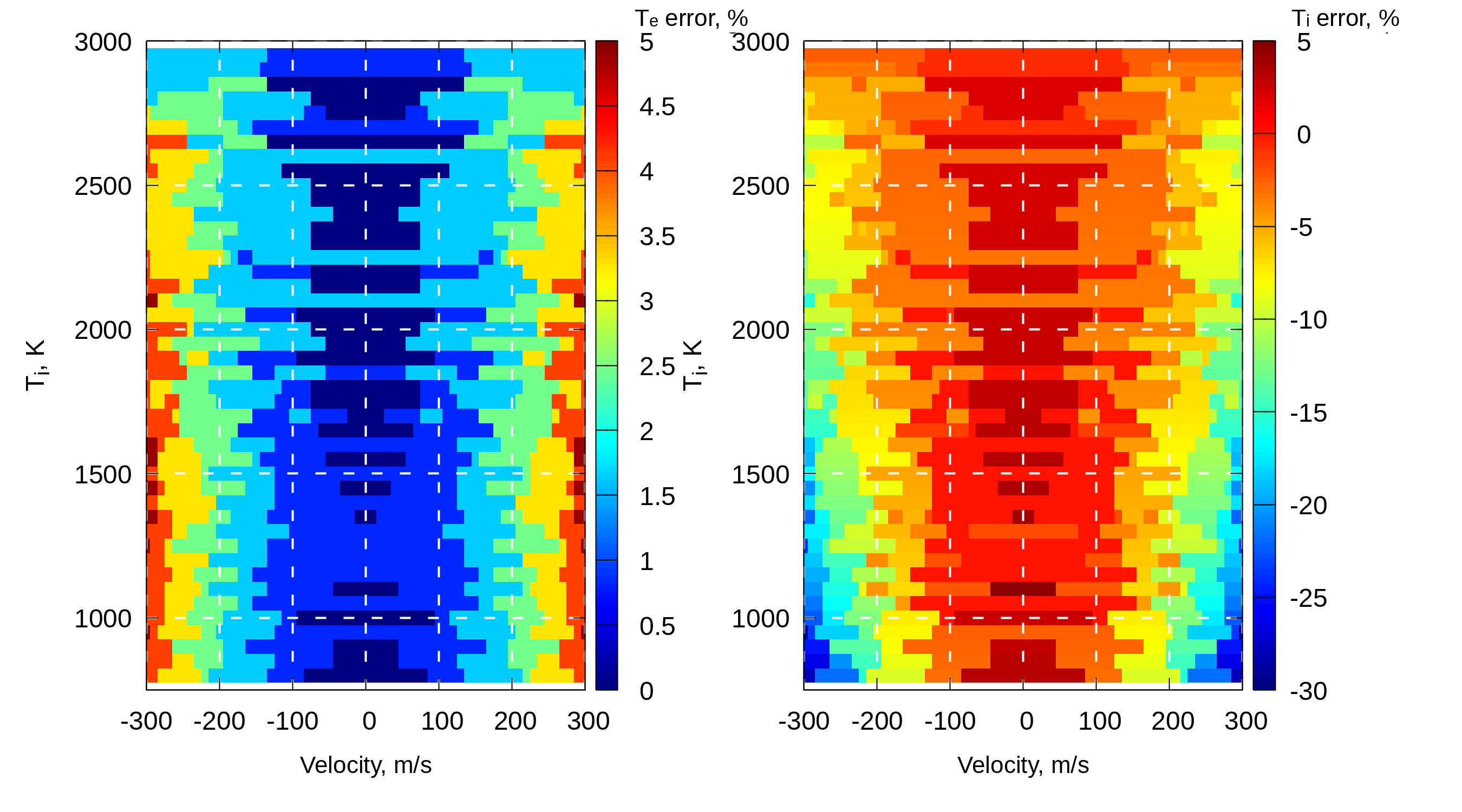}
\caption{Dependence of $T_e$ and $T_i$ recovery error on drift velocity $V_d$ and ion temperature $T_i$ for a fixed electron temperature $T_e = 3000$~K.}
\label{fig:Err_image}
\end{figure}

\newpage
\begin{figure}[h!]
\centering
\includegraphics[width=0.8\textwidth]{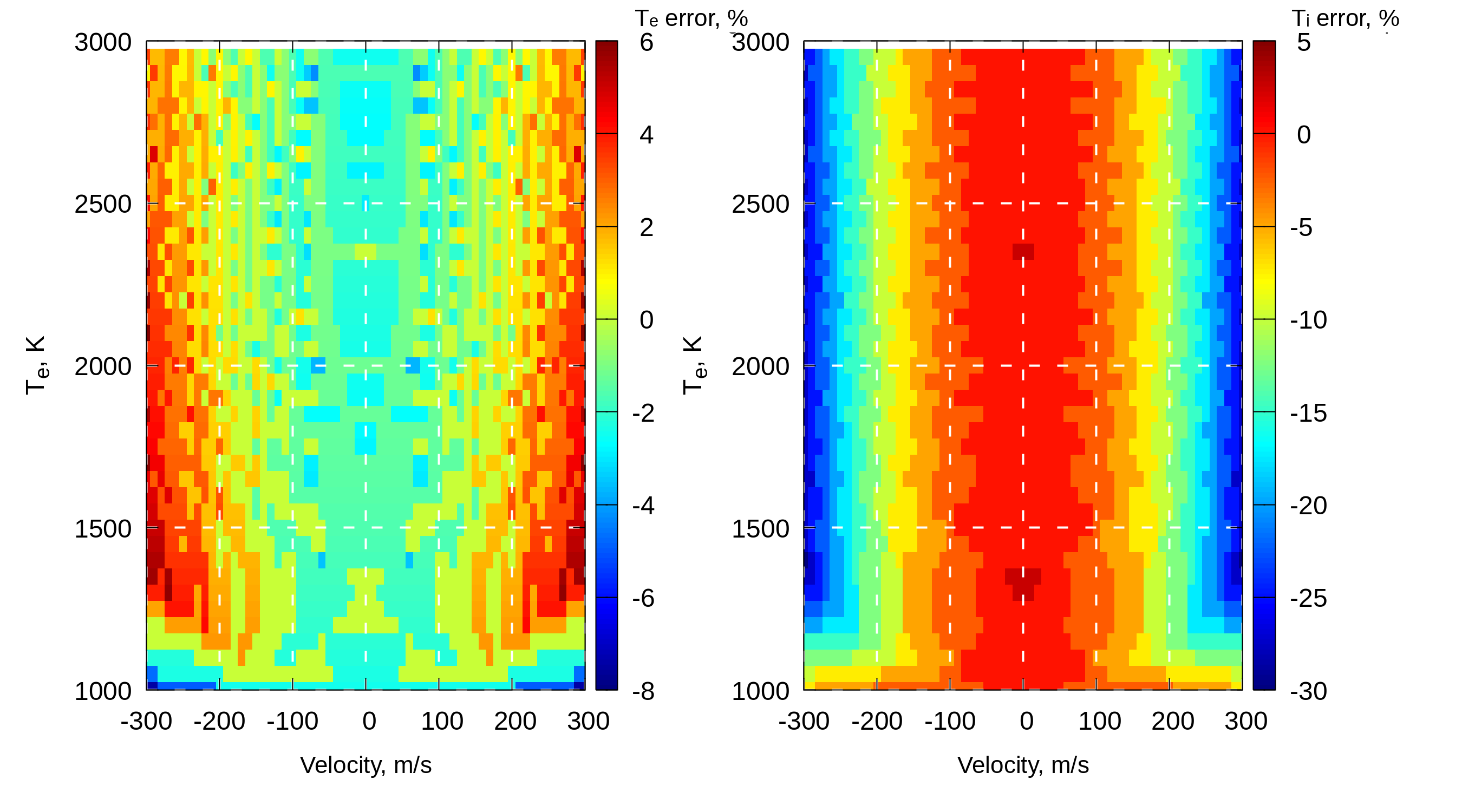}
\caption{Dependence of $T_e$ and $T_i$ recovery error on drift velocity $V_d$ and electron temperature $T_e$ for a fixed ion temperature $T_i = 1000$~K.}
\label{fig:Err_image_Te}
\end{figure}

\newpage
\begin{figure}[h!]
\centering
\includegraphics[width=0.8\textwidth]{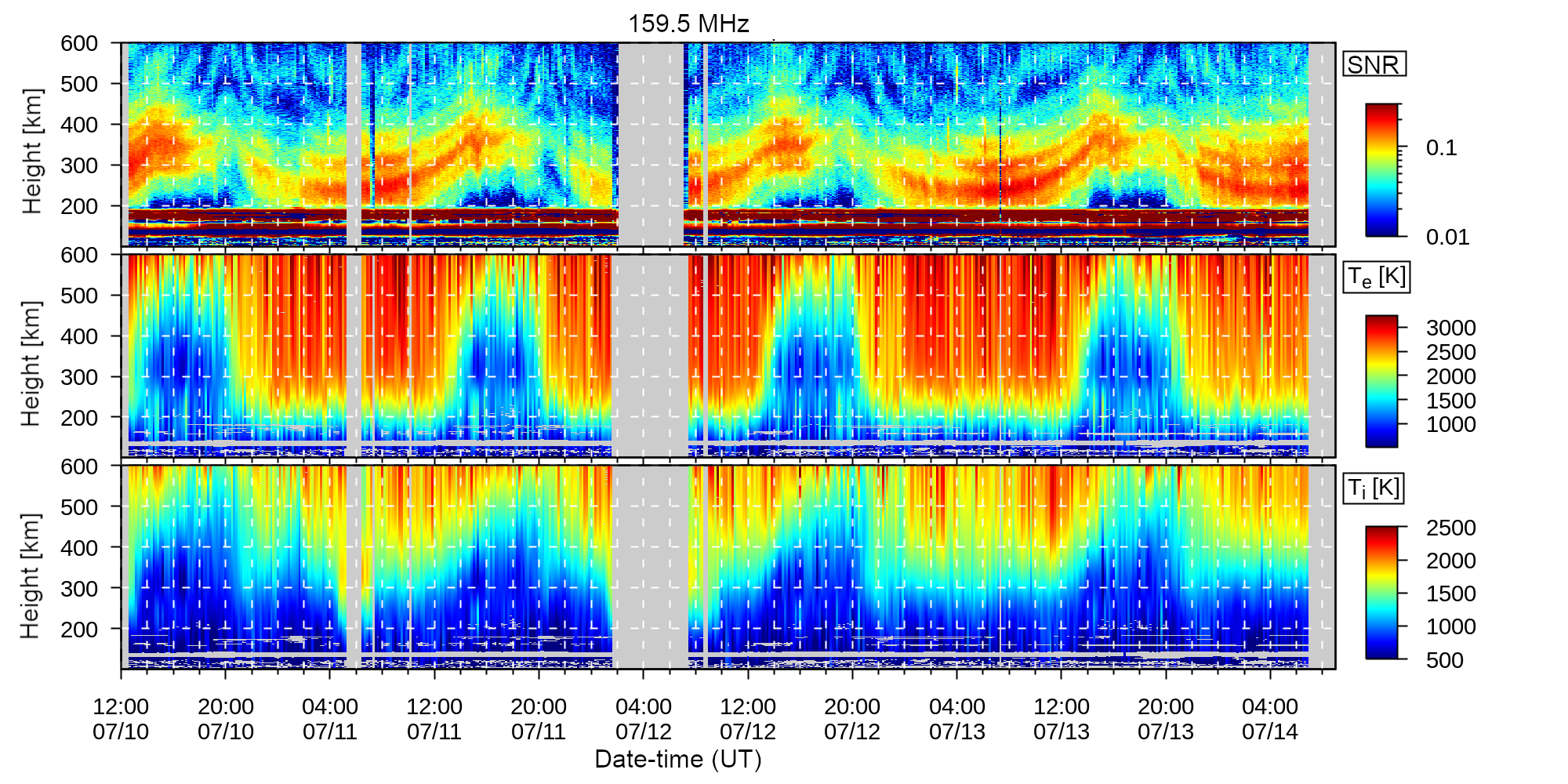}
\caption{Results of narrowband signal analysis (Channel 0, frequency 159.5~MHz, accumulation time 10 minutes) with the additional constraint enabled (equation \eqref{eq:anchor_inequality}). Top panel: signal-to-noise ratio. Middle panel: electron temperature. Bottom panel: ion temperature.}
\label{fig:res_image}
\end{figure}

\newpage
\begin{figure}[h!]
\centering
\includegraphics[width=0.8\textwidth]{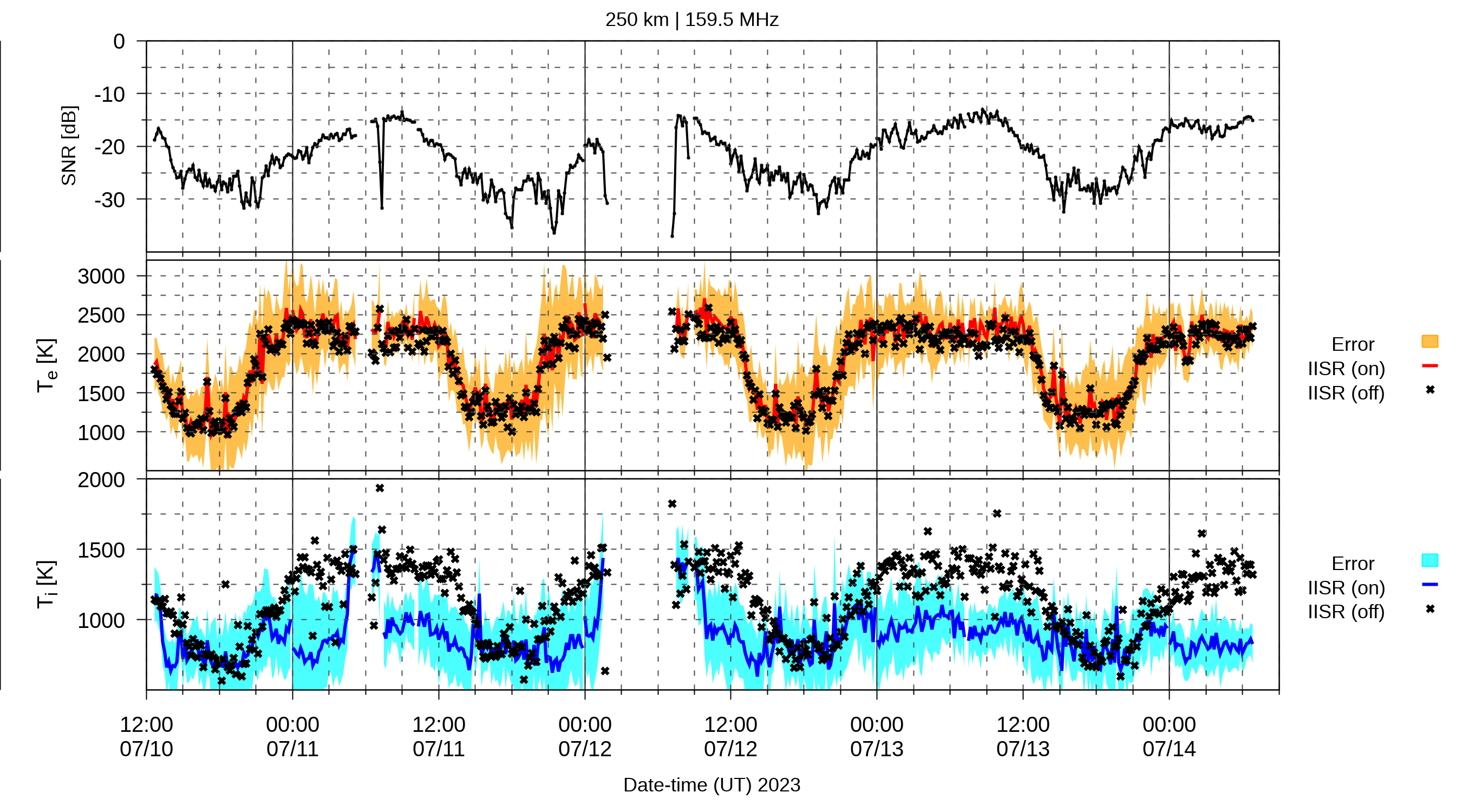}
\caption{Comparison of data with the additional constraint enabled (red line for $T_e$, blue line for $T_i$) and without it (black dots) for an altitude of 250~km. Temperature error bounds for the constrained case are shown as shaded areas (orange for $T_e$, light blue for $T_i$). Top panel: signal-to-noise ratio. Middle panel: electron temperature. Bottom panel: ion temperature.}
\label{fig:res_250}
\end{figure}

\newpage
\begin{figure}[h!]
\centering
\includegraphics[width=0.8\textwidth]{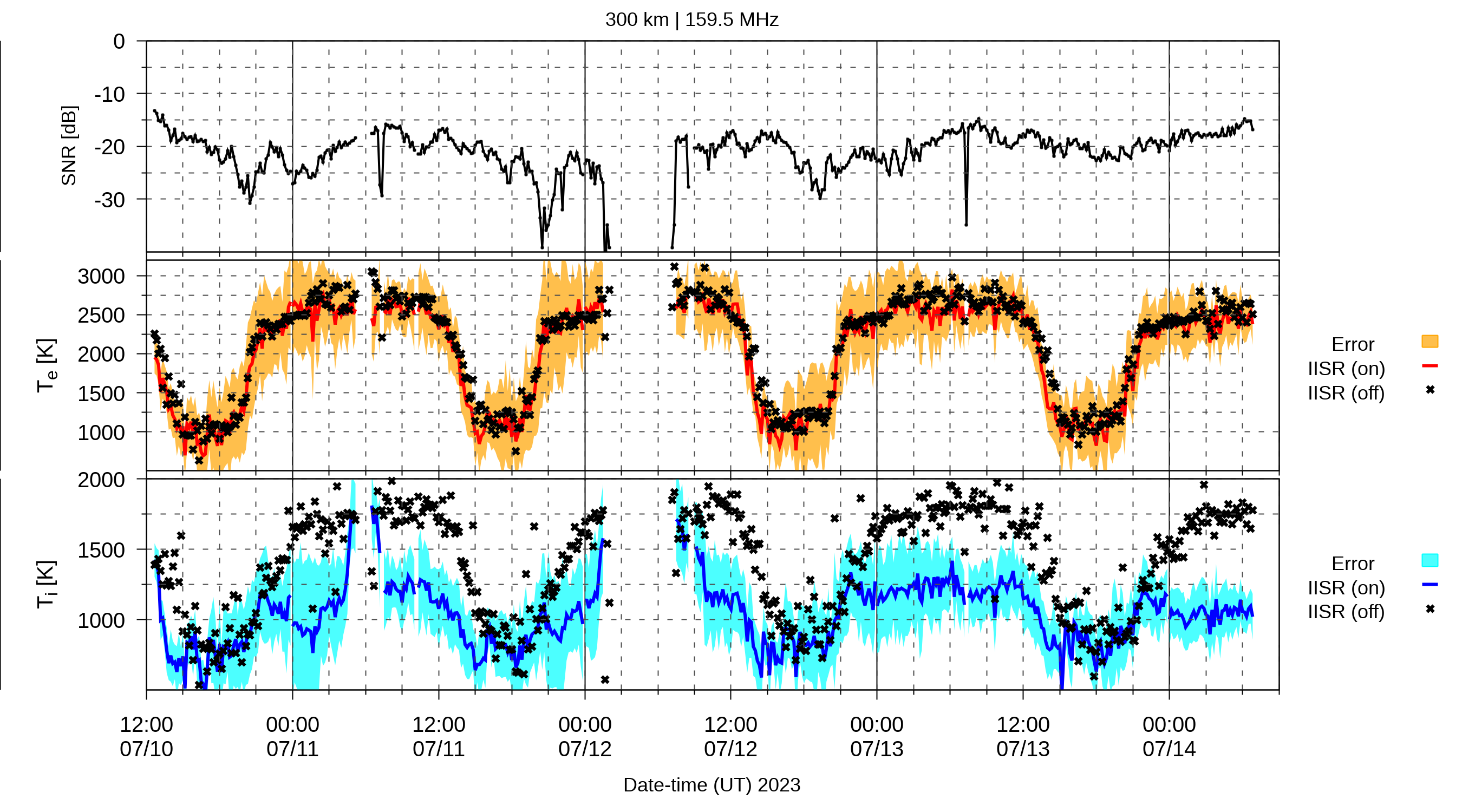}
\caption{Comparison of data with the additional constraint enabled (red line for $T_e$, blue line for $T_i$) and without it (black dots) for an altitude of 300~km. Temperature error bounds for the constrained case are shown as shaded areas (orange for $T_e$, light blue for $T_i$). Top panel: signal-to-noise ratio. Middle panel: electron temperature. Bottom panel: ion temperature.}
\label{fig:res_300}
\end{figure}

\newpage
\begin{figure}[h!]
\centering
\includegraphics[width=0.8\textwidth]{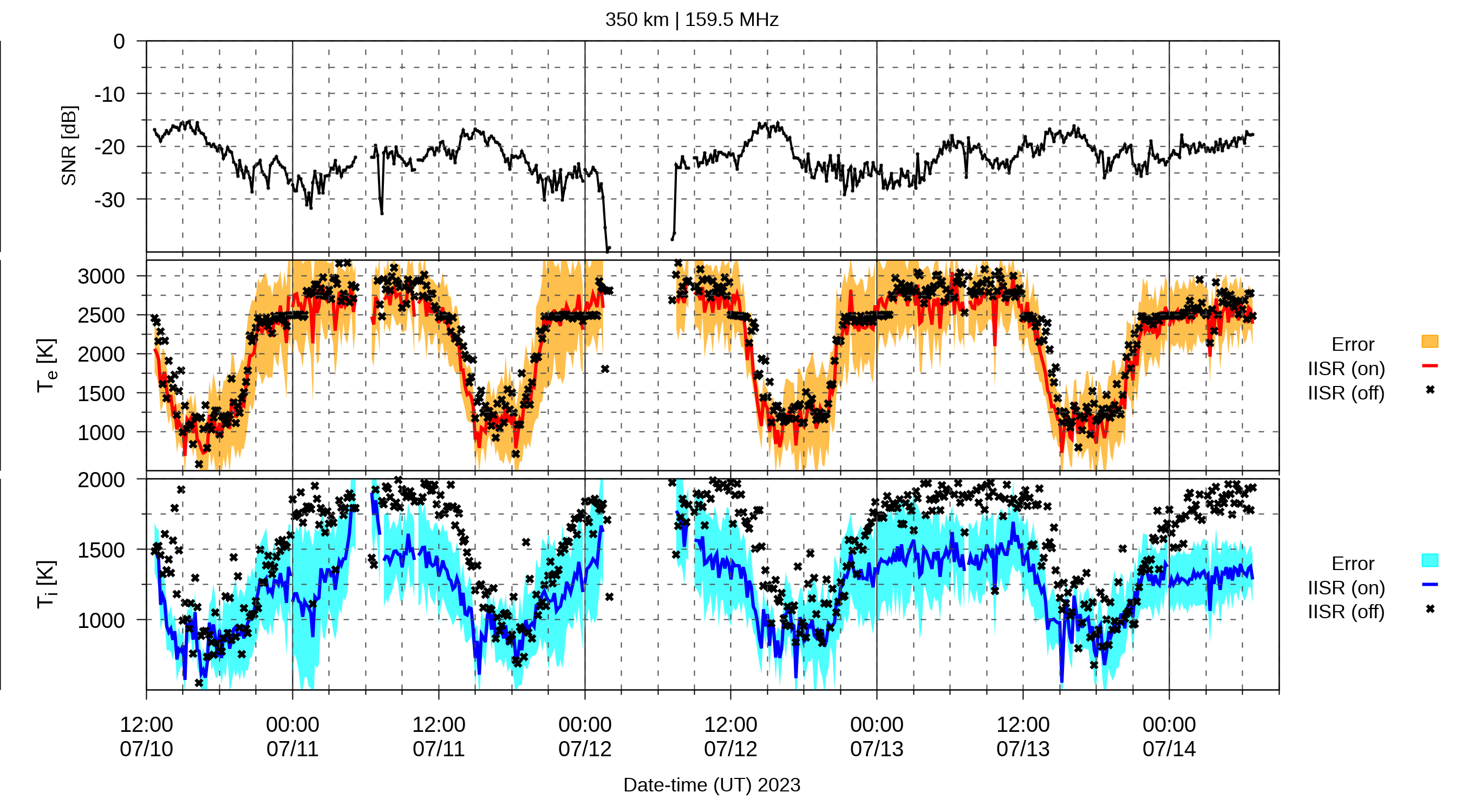}
\caption{Comparison of data with the additional constraint enabled (red line for $T_e$, blue line for $T_i$) and without it (black dots) for an altitude of 350~km. Temperature error bounds for the constrained case are shown as shaded areas (orange for $T_e$, light blue for $T_i$). Top panel: signal-to-noise ratio. Middle panel: electron temperature. Bottom panel: ion temperature.}
\label{fig:res_350}
\end{figure}

\newpage
\begin{figure}[h!]
\centering
\includegraphics[width=0.8\textwidth]{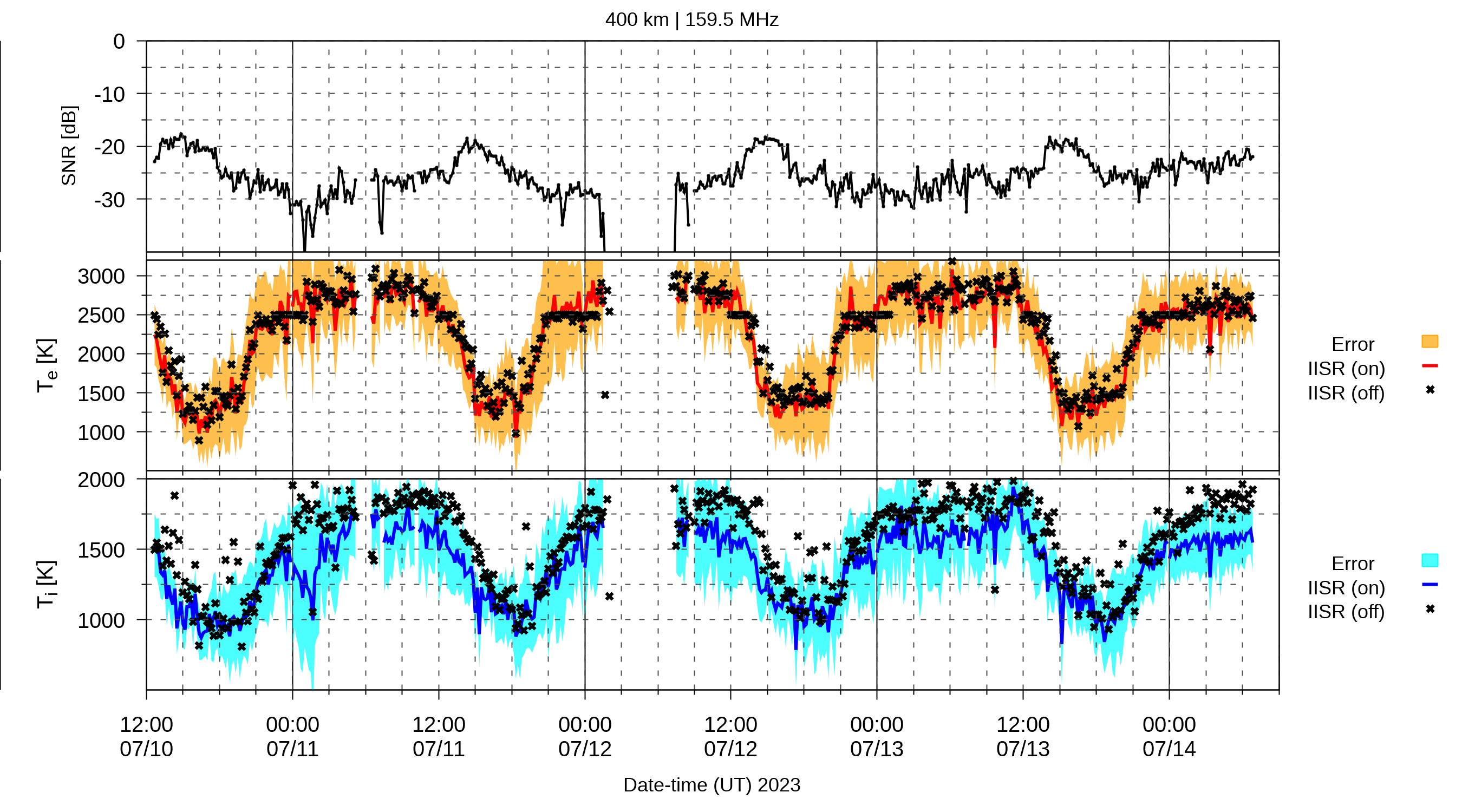}
\caption{Comparison of data with the additional constraint enabled (red line for $T_e$, blue line for $T_i$) and without it (black dots) for an altitude of 400~km. Temperature error bounds for the constrained case are shown as shaded areas (orange for $T_e$, light blue for $T_i$). Top panel: signal-to-noise ratio. Middle panel: electron temperature. Bottom panel: ion temperature.}
\label{fig:res_400}
\end{figure}

\end{document}